\documentstyle[epsfig]{mn}
\def\spose#1{\hbox to 0pt{#1\hss}}
\def\lsim{\mathrel{\spose{\lower 3.0pt\hbox{$\mathchar"218$}}
     \raise 2.0pt\hbox{$\mathchar"13C$}}}
\def\gsim{\mathrel{\spose{\lower 3.0pt\hbox{$\mathchar"218$}}
     \raise 2.0pt\hbox{$\mathchar"13E$}}}
\newcommand{\Aa}{{\it Aa}}
\newcommand{\Ae}{{\it Ae}}
\newcommand{\Ie}{{\it Ie}}

\def\msun{{\rm\,M_\odot}}

\def\spose#1{\hbox to 0pt{#1\hss}}
\def\lta{\mathrel{\spose{\lower 3pt\hbox{$\mathchar"218$}}
     \raise 2.0pt\hbox{$\mathchar"13C$}}}
\def\gta{\mathrel{\spose{\lower 3pt\hbox{$\mathchar"218$}}
     \raise 2.0pt\hbox{$\mathchar"13E$}}}

\title[Evolution of Multi-mass Globular Clusters]
{Evolution of Multi-mass Globular Clusters in Galactic Tidal 
Field with the Effects of Velocity Anisotropy}
\author[Takahashi and Lee]
{K.~Takahashi$^1$\thanks{e-mail: takahasi@chianti.c.u-tokyo.ac.jp 
}, H. M.~Lee$^2$\thanks{e-mail: hmlee@astro.snu.ac.kr} \\
$^1$ Department of Astronomy,
School of Sciences, The University of Tokyo,
7-3-1 Hongo, Bunkyo-ku, Tokyo 113-0033, Japan\\
$^2$Department of Astronomy, Seoul National University, 
Seoul 151-742, Korea}

\date{ Received 1999 September 00}
\pagerange{\pageref{firstpage}--\pageref{lastpage}}
\pubyear{2000}

\begin{document}
\maketitle

\label{firstpage}
\begin{abstract}
We study the evolution of globular clusters with mass spectra
under the influence of the steady Galactic tidal field, including the effects 
of velocity anisotropy. Similar to
single-mass models, velocity anisotropy develops as the cluster evolves,
but the degree of anisotropy is much smaller than isolated clusters.
Except for very early epochs of the cluster evolution,
nearly all mass components become tangentially anisotropic at the outer parts.
We have compared our results with 
multi-mass, King-Michie models. The isotropic King model better fits to the 
Fokker-Planck results because of tangential anisotropy. 
However, it is almost impossible to fit the computed density profiles
to the multi-mass King 
models for all mass components. Thus if one attempts to derive global mass 
function based on the observed mass function in limited radial range using 
multi-mass King models, one may get somewhat erratic results,
especially for low mass stars. 
We have examined how the mass function changes in time. 
Specifically, we find that the power-law index of the mass function 
decreases monotonically
with the total mass of the cluster. This appears to be consistent with 
the behaviour of the observed slopes of mass functions for a limited number of 
clusters, although it is premature to compare quantitatively because there
are other mechanisms in contributing the evaporation of stars from the
clusters. 
The projected velocity profiles for anisotropic models with  the
apocenter criterion for evaporation show significant flattening toward
the tidal radius compared to isotropic model or anisotropic model with the
energy criterion. Such a behaviour of velocity profile appears to be 
consistent with the observed profiles of collapsed cluster M15.
\end{abstract}


\section{INTRODUCTION}

The study of evolution of globular clusters has a long history. It would 
be most desirable to use direct $N$-body calculations in order to follow 
the evolution of realistic globular clusters realistically, 
but we do not have enough 
computing power to do that for $N$ comparable to the number
of stars in real globular clusters at this moment. 
Therefore 
we have to rely on approximate techniques to understand the dynamics of 
star clusters. 
The Fokker-Planck equation has been widely used for the study of globular 
clusters, but it requires many simplifying assumptions. 
The assumption of velocity 
isotropy is usually employed because it allows fast and accurate 
integration of the Fokker-Planck equation (Cohn 1980). However, high 
degree of velocity anisotropy is known to be generated by two-body 
relaxation in the outer parts of isolated clusters (e.g., Spitzer 1987). The 
observational properties of the clusters depend on the degree of anisotropy
and correct treatment of anisotropy becomes an important issue. 
The development of velocity anisotropy was investigated
in some recent works which used various numerical methods:
e.g., direct $N$-body calculations by Giersz \& Heggie (1994), 
gaseous model calculations by Louis \& Spurzem (1991) and Spurzem (1996), 
and Fokker-Planck calculations by Cohn (1979), Drukier et al. (1999) and
Takahashi (1995, 1996, 1997). Most of these calculations confirmed the
earlier expectations of generation of radial anisotropy in the outer
parts of the clusters, although detailed behaviour depends somewhat on the
numerical methods.

Actual clusters are embedded in the Galactic tidal field which should
impose a finite boundary to the cluster. The introduction of tidal boundaries 
should have significant effects on the velocity anisotropy.
Recently Takahashi, Lee \& Inagaki (1997; abbreviated as TLI hereafter) 
performed anisotropic Fokker-Planck simulations for single mass
clusters and
demonstrated that radial 
anisotropy grows up to the time of core collapse,
but the degree of anisotropy begins to decrease to 
smaller values as the cluster expands and loses significant 
amount of mass. They showed 
that there is even {\it tangential} anisotropy very close to the tidal 
boundary in the late phase of the cluster's evolution. The density 
profiles of the clusters can be reasonably fitted by isotropic King models 
for most of the time except for the brief phase around the core collapse, when
the degree of radial anisotropy nearly reaches its maximum.

The behaviour of the anisotropy for isolated clusters with mass 
spectra has some interesting features. Takahashi (1997) showed that the 
degree of anisotropy 
for different mass components behaves 
differently. There is always radial anisotropy for low mass components, 
but tangential anisotropy can develop for high mass components
at early evolutionary stages when
the temperature of heavy stars decreases rapidly toward the
equipartition. 
During this `cooling' phase, the radial
velocity dispersion decreases more rapidly than the
tangential one. This leads to the development of tangential anisotropy
among high mass stars.

As in the case of single mass models, extension of these studies into tidally
limited clusters should provide useful insight on the evolution of realistic
clusters. Based on the experience of single mass models, one obviously
expects that the velocity anisotropy will be greatly affected by the
presence of the tidal field. Some of the essential points have been
clarified by the analyses of $N$-body results (Vesperini \& Heggie 1997), but
one should remember that the $N$-body calculations are done for small $N$ cases.
Many of the behaviour of self-gravitating systems does not allow simple
scaling (e.g., Portegies Zwart et al. 1998, Takahashi \& Portegies Zwart 1998). 
We make detailed calculations for the evolution of 
globular clusters with these effects in order to make quantitative 
assessments for the long term evolution of star clusters with the tidal cut-off
and mass spectrum.

The study of 
the evolution of tidally limited clusters has important areas of 
application. The mass function of globular clusters is difficult to 
obtain because of large dynamic range in brightness of 
individual stars as well as large variation of stellar densities. 
It is very difficult 
to determine the {\it global} mass function observationally, but there 
have been several 
studies to determine the global mass function based on {\it locally} measured 
mass functions in limited areas of 
globular clusters (Richer et al. 1990, 1991 and references therein).  
It is absolute necessity 
to obtain the global mass function if one wants to study factors 
determining the mass function of clusters. The conversion of the local mass 
function to the global one is not a trivial task. The general practice 
employed widely is to fit the observation to multi-mass King models
(e.g., Meylan 1988) or to
assume that the mass function in low mass stars in the outer parts are
close to the global one (Richer et al. 1991). 
However, there is no guarantee that these processes give a right answer. 
The detailed studies of evolving clusters with the tidal boundary can provide
an important check of the above mentioned practices.

The mass function changes with time because the rate of escape of stars 
depends on the mass of the star. The high mass stars tend to spiral into 
the central parts by dynamical friction, and consequently low mass stars are 
preferentially removed from the tidally limited clusters. The slope of 
mass function thus becomes less steep as the cluster loses mass. Even 
if one gets the global mass function through careful observations and
correct conversion process, the present day mass 
function (PDMF) may be significantly different from the initial mass 
function (IMF). The detailed studies of evolving star clusters 
could provide an important framework of estimating the relation between
PDMF and IMF.

In the present paper, we make detailed studies of the the evolution of
tidally limited multi-mass clusters including the effects of velocity 
anisotropy. We pay special attention to the behaviour of mass 
function during the course of dynamical evolution.  We examine 
the adequacy of the process of converting the local mass function to the
global mass function using multi-mass King model fitting.
We then analyze how the mass 
function changes with time and location. 

This paper is organized as follows. 
In \S\,\ref{sec:FPmodels}, we describe our models for 
the evolution of globular clusters. 
In \S\,\ref{sec:ch}, we introduce the idea of half-mass concentration
and discuss its relation to the development of anisotropy and to
mass loss rates.
In \S\,\ref{sec:MMmodels} we describe initial conditions of 
multi-mass models.
In \S\,\ref{sec:results}  we present the results of the 
model calculations. The relationship between local and global mass function 
and the temporal evolution of mass function are
examined in \S\,\ref{sec:mf}. We discuss the velocity dispersion
profiles in \S\, \ref{sec:vd}. The final section summarizes basic results.

\section{Fokker-Planck Models}\label{sec:FPmodels}

We use the anisotropic Fokker-Planck code developed by Takahashi (1995).  
It is based on the direct integration method of the Fokker-Planck equation 
in $(E,J)$ space (Cohn 1979),
where $E=v^2/2 +\phi (r)$ is the energy of a star per unit mass 
and $J=rv_t$ is the angular momentum per unit mass. The extension to 
multi-mass clusters is described in Takahashi (1997) and the treatment of 
tidal boundary can be found from TLI. TLI used two different criteria for 
the escape of a star from the cluster: apocenter criterion and energy 
criterion. The {\it apocenter} criterion assumes that a star escapes if 
the apocenter distance of a star with $(E,J)$ becomes greater than tidal 
radius $r_t$. This criterion preferentially removes the stars with low 
$J$ and thus suppresses the development of radial anisotropy.  The
$energy$
criterion assumes that a star escapes if $E>E_t$ where $E_t = \phi 
(r_t)$. The radial anisotropy is also suppressed in this criterion and 
tangential anisotropy can develop (see TLI).
The general behaviour of 
the cluster evolution is not much dependent on the choice of escape 
criteria for the models of TLI, 
but details of the clusters vary with the treatment of 
stellar evaporation. Takahashi \& Portegies Zwart (1998) have shown that
the apocenter criterion gives better results when compared with
realistic $N$-body calculations. In the present study,
we examine the results of our calculations with both 
of the criteria, but more emphasis is given to the calculations
with the apocenter criterion.

Hereafter, we denote anisotropic models with the apocenter criterion as
{\it Aa}
models and anisotropic models with the energy criterion as {\it Ae}
models.
The capital {\it A} stands for anisotropic models, and the lower cases
{\it a}
and {\it e} stand for apocenter and energy criteria, respectively.
Similarly isotropic models are denoted as {\it Ie} models,
but {\it Ia} models do not exist.

\section{Half-Mass Concentration,
Velocity Anisotropy, and the Mass Loss Rate}\label{sec:ch}

Before going into descriptions of the results of multi-mass models,
we examine the effects of the degree of
the concentration of a cluster on the development of velocity anisotropy
and on the cluster mass loss rate.
We use single-mass models in this section
because these models are simple, but
the conclusions of this section are also true for multi-mass models,
as we show later.

TLI used the initial model of a single-mass Plummer model 
with the cutoff radius of 32 in Plummer model units.
The results of their simulations show that
the \Aa\ and \Ae\ models lose mass faster than the \Ie\ model
(see Fig. 2a of TLI, though the \Ae\ model
is not shown in it).
The evaporation time of the \Ie\ model
is about 12\% longer than that of the \Aa\ model,
and the latter is about 10\% longer
than that of the \Ae\ model.
The biggest difference in the mass loss rate
between the anisotropic models and the isotropic
model appears in the pre-collapse and the early post-collapse phases.
The mass loss rates of the different models are similar
in the late post-collapse phase.

In \S\,\ref{sec:results} we show the results of simulations
which start with multi-mass King models.
As shown in Fig. \ref{massfig}, the \Ie\ model and the \Ae\
model are very similar in the evolution of the total mass.
The \Aa\ model loses mass
{\it slower} than the \Ie\ model.
This trend is opposite to that seen in the simulations of TLI.
While Fig. \ref{massfig} shows the results for a $W_0=3$ King model,
the same trend is seen for King models with other $W_0$.
Why does the opposite trend appear?
The change from the single-mass model to the multi-mass model
is not the reason of the change of the trend.
Rather, it is found that the trend changes with the 
concentration of a cluster to the tidal cutoff radius.

First we discuss what parameter is useful in measuring
the degree of the concentration of a cluster.
For King models, the concentration
is usually measured in terms of the concentration parameter
$c \equiv \log_{10} (r_{\rm t}/r_0)$, where $r_0$ is the King core
radius
(King 1966).
The parameter $c$ is useful in measuring the concentration of the core.
However, when discussing the variation of the total mass, $c$ is not
always an appropriate parameter, because the fraction of the core mass
is
very small in large $c$ clusters.
It may be more appropriate to use
the half-mass radius $r_{\rm h}$ instead of the core radius $r_0$.
Thus we define half-mass concentration
\begin{equation}
c_{\rm h} \equiv r_{\rm t}/r_{\rm h}, \label{eq:ch}
\end{equation}
(note the logarithm is not taken here).

\begin{figure}
\centerline{\epsfig{figure=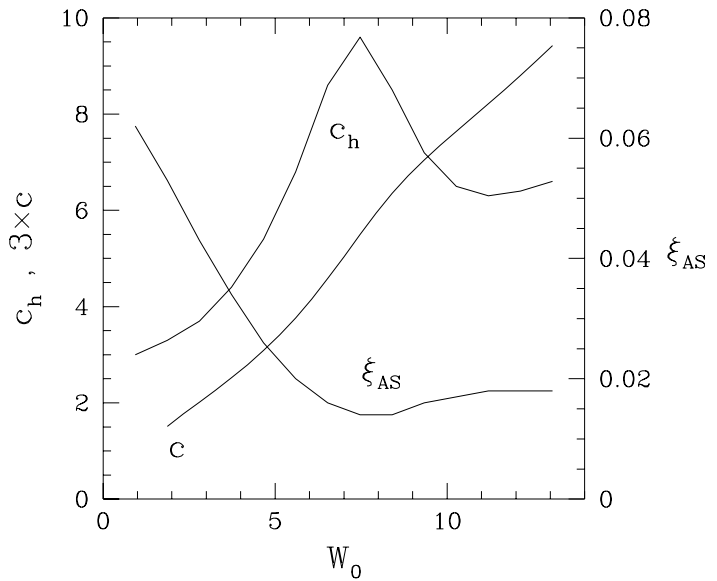, height=10cm}}
\caption{The behaviour of $c_h$, $c$,  and $\xi_{AS}$ for King models
against $W_0$. Note that $c_h$ has a maximum, and $\xi_{AS}$ has
a broad minimum near $W_0 \simeq 8$.}
\label{fig:king}
\end{figure}

King models form a one-parameter family parameterized by
the concentration $c$ or equivalently by
the dimensionless central potential depth $W_0$;
$c$ monotonically increases with $W_0$ (King 1966).
On the other hand, $c_{\rm h}$ does vary monotonically with
$W_0$.
The relation among $W_0$, $c$, and $c_{\rm h}$
is given in Fig. \ref{fig:king} for $1 \le W_0 \le 14$.
In contrast to that
the core concentration $c$ varies by nearly three orders of magnitude,
the variation of the half-mass concentration $c_{\rm h}$ is
within a factor of three
for this range of $W_0$.
The half-mass concentration $c_{\rm h}$ increases with $W_0$
for $W_0=1$ to 8, then it decreases until $W_0$ reaches  and increases
again.
For the range of $1 \le W_0 \le 14$,
$c_{\rm h}$ reaches the maximum of about 10 at $W_0 \simeq 8$.
The initial model of TLI has $c_{\rm h}=25$,
therefore it is far more concentrated than
any King models in terms of the half-mass concentration.

\begin{figure}
\centerline{\epsfig{figure=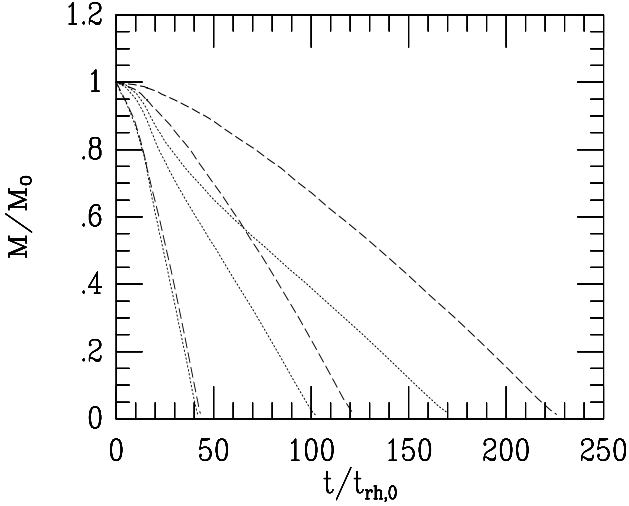, height=7cm,width=9cm}}
\caption{
The evolution of mass as a function of time
for \Ae\ models (dotted lines) and \Ie\ models (dashed lines).
The initial models are single-mass Plummer models with three different
tidal cutoff radii: $c_{\rm h} \equiv r_{\rm t}/r_{\rm h}=$10, 20, 30,
from left to right.
As $c_{\rm h}$ decreases the mass loss becomes faster and
the difference between the \Ae\ model and the \Ie\ model becomes
smaller.
The core collapse occurs at $t=$16--18$t_{\rm rh,0}$
in all the models.
}
\label{fig:plmass}
\end{figure}

\begin{figure}
\centerline{\epsfig{figure=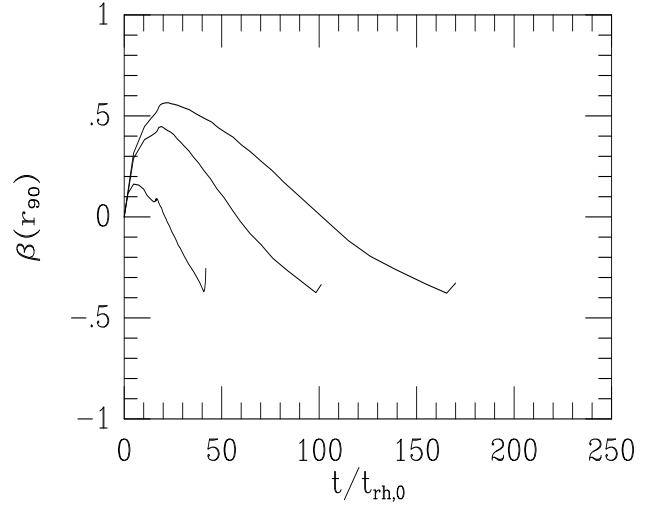, height=7cm,width=9cm}}
\caption{
The evolution of the anisotropy $\beta$ at the 90\%-mass radius
for the \Ae\ models plotted in Fig. \ref{fig:plmass}.
The initial development of the radial anisotropy
becomes strongly suppressed as $c_{\rm h}$ decreases from 30 to 10.
}
\label{fig:plbeta}
\end{figure}

In order to investigate
the effects of the half-mass concentration on the cluster evolution,
we carry out simulations which start with single-mass Plummer models
with different cutoff radii
which correspond to $c_{\rm h}=$10, 20, and 30.
In these simulations
the number of particle is set to $10^4$, which is specified
to include the effects of three-body binary heating.
Fig. \ref{fig:plmass} shows the evolution of the total mass
for the \Ae\ and \Ie\ models.
For the large $c_{\rm h}$ initial models,
the \Ae\ model
loses mass much faster than the \Ie\ model at the early evolutionary
phase as in TLI's simulations.
(The core collapse occurs at $t=$16--18$~t_{\rm rh,0}$
in all the models.)
As $c_{\rm h}$ decreases, the difference between the \Ae\ and
\Ie\ models decreases
and finally almost diminishes for $c_{\rm h}=10$.
Fig. \ref{fig:plbeta} shows the evolution of the anisotropy $\beta$
at the 90\%-mass radius.
The anisotropy
parameter $\beta$ is defined as follows:
\begin{equation}
\beta \equiv 1-{\sigma_t^2\over \sigma_r^2},
\label{eqn:beta}
\end{equation}
where $\sigma_t$ and $\sigma_r$ are one-dimensional tangential and radial
velocity dispersions, respectively.
The development of the radial anisotropy during the pre-collapse phase
becomes strongly suppressed as $c_{\rm h}$ decreases.
This is simply because radial-orbit stars can easily escape
from a small tidal radius cluster.
As TLI discussed, the emergence of a large number of radial-orbit stars
is the reason for faster mass loss in the anisotropic models.
For low $c_{\rm h}$ clusters,
since there is almost no room for the radial anisotropy
to develop,
the \Ae\ model and the \Ie\ model do not differ very much
in the evolution of the total mass.
Since King models have such low $c_{\rm h}$ ($c_{\rm h}<10$),
the difference among the three models
shown in Fig. \ref{massfig} can be understood from this viewpoint.
The mass loss of the \Aa\ model should be always slower than that of
the \Ae\ model because of the apocenter criterion (see TLI).

Note that we have assumed that the initial tidal radius is equal to
the limiting radius of King models.
This is a conventional way for setting up initial conditions
of simulations, but not the only one;
these two radii can be different.

Finally in this section,
we comment on the relation between the mass loss rate
and the half-mass concentration for King models.
Johnstone (1993, figure 2) and
Gnedin \& Ostriker (1997, figure 6)
found that for King models the mass loss rate per half-mass relaxation
time (exactly speaking, the inverse of the evaporation time in units
of the initial half-mass relaxation time in Gnedin \& Ostriker)
has a minimum around $c=$1.5--2.
Johnstone (1993) calculated the evaporation rate for King models
using the local Fokker-Planck equation,
and Gnedin \& Ostriker (1997) performed isotropic orbit-averaged
Fokker-Planck
simulations with the initial conditions of King models.
Their findings are qualitatively consistent with a simple estimation
of the mass loss rate
based on only the half-mass concentration $c_{\rm h}$.
Naively one may think that the mass loss rate decreases as $c_{\rm h}$ 
increases. In fact
$c_{\rm h}$ has the maximum at $W_0 \simeq 8$ ($c \simeq 1.8$),
where Johnstone (1993) and Gnedin \& Ostriker (1997) found the
minimum evaporation rate.
In addition
Johnstone (1993) found a local maximum of the mass loss rate
at $c \simeq 2.6$. This may correspond to the fact that
$c_{\rm h}$ has a local minimum at $c \simeq 2.7$ ($W_0 \simeq 12$)
as shown in Fig. \ref{fig:king}.
As an additional reference,
we show the mass loss rate per half-mass relaxation time
$\xi_{\rm AS}$
calculated from the modified Ambartsumian-Spitzer formula
(Spitzer 1987, p.57; see also TLI) for King models in Fig.
\ref{fig:king}.
This estimation is even quantitatively not very different from the
result of more detailed calculations for $W_0 \le 8$ by Johnstone
(1993).

\section{Multi-Mass Models}\label{sec:MMmodels}

There are growing evidences that the mass function in clusters as well as
in the field is not a simple power-law, but should be approximated as 
composite power-laws (e.g., Kroupa, Tout \& Gilmore 1993). However,
we use simple power-laws in the present study of general evolution of
clusters, for simplicity. The adoption of multiple composite power-laws
is trivial task, but requires specification of more model parameters.
The assumed initial mass function (IMF) has the following form:
\begin{equation}
N(m) dm = C m^{-\alpha} dm, \qquad m_{\rm min}\le m \le m_{\rm max},
\label{eq:pmf}
\end{equation}
where $C$ and $\alpha$ are constants. 
As in Takahashi (1997), we use the discrete mass components. 
The $i$-th mass bin for equal $\Delta \log m$ interval can be obtained by
\begin{equation}
m_i = m_{\rm min} \left({m_{\rm max}\over m_{\rm min}}\right)^{(i-{1\over 2})/K}, 
\quad (i=1, 2, ..., K),
\end{equation}
where $K$ is the number of components. The parameters determining the 
mass function are $\alpha$ and $\mu (\equiv {m_{\rm max}/m_{\rm min}})$. We use 
a few values of $\alpha$ and $\mu$ for our model calculations.

The issue of the effects of stellar evolution naturally comes out for 
multi-mass models if one extends the mass function to high mass stars. 
Obviously the stellar evolution should have been the most important 
process in determining the early phase of the 
dynamical evolution of globular clusters. 
However, we will ignore the stellar evolution in this paper
because it adds one more complexity to the models. 

Without stellar evolution, the clusters usually undergo core-collapse 
before losing significant amount of stars unless the initial cluster has 
very low central concentration, because the time scale for the core collapse
is generally much shorter than the evaporation time. Note that the
core collapse time is of order of $10 t_{\rm rh,0}$ while the evaporation time 
is typically of order of $40 t_{\rm rh,0}$, where $t_{\rm rh,0}$ is the 
initial half-mass relaxation time for single mass models. 
The core collapse is very much accelerated for
multi-mass clusters, and the discrepancy between the evaporation time 
and the core collapse time could become even larger. 
Therefore one needs to supply the mechanism
that allows the evolution of cluster beyond the core collapse. 
We assumed that the heating is provided by binaries 
formed by three-body processes because it is easiest to implement in the
framework of Fokker-Planck method. 
The evolution of parameters for the central parts 
such as density and velocity dispersion should depend on the detailed mechanism 
for driving the post core-collapse evolution, but most of the global 
properties do not sensitively depend on the heating mechanism. 
For our purposes, 
three-body binary heating is sufficiently simple and easy to be used as 
heating mechanism. The heating formula for multi-mass clusters is taken 
from Lee, Fahlman \& Richer (1991). 

\section{Results of Multi-Mass Model Calculations}\label{sec:results}

\subsection{Initial Model Parameters}
We have used various initial models. The characteristics of the
initial models are summarized in Table 1. 
In all models, we have assumed that the shape of density 
profiles of different mass components are the same with the different 
amplitudes dictated by the assumed mass functions. The velocity dispersions
are assumed to be the same for all components.
are assumed to have the same density and velocity profiles. 

The behaviour of the cluster up to the core collapse does not depend on
the initial total mass or size of the cluster. But the amount of binary heating
depends on the number of stars in the cluster. We have assumed that the
mass of the cluster is $6 \times 10^4 ~ \msun$ for
the model $R$, and we have fixed the total number of stars at 
$5\times 10^4$ for other models. But this number 
only controls the central density during the postcollapse phase. 
Most of our results does not really depend on the mass of the cluster,
except for the surface density profiles. If we used higher initial mass,
the core radii at a given epoch (measured by the $M/M_0$) would be smaller.
The minimum mass $m_{min}$ of the mass function was assumed to be 
0.1 $\msun$ in all models.

Throughout this paper, we use $r$ to represent the distance from the
center of the cluster, and $R$ to represent the {\it projected} distance.

\subsection{A Reference Model}\label{sec:heggie}

We first describe the general behaviour of {\it Reference Model} (Model R)
whose initial parameters are given in Table 1. This set of parameters is used
by Heggie et al. (1998) to compare the outcomes of various 
computational methods. We will consider this initial condition as a reference
model. The number of mass component was 8.

\begin{table*}
\centering
\begin{minipage}{140mm}
\caption{Initial Parameters for Multi-Mass Models}
\begin{tabular}{llccl}
\hline
Model & Profile & $\alpha$ & $\mu$ & Comments \\
\hline
R & King, $W_0=3$ & 2.35 & 15 & reference \\
K1 & King, $W_0=3$ & 2.35 & 7 &  low $\mu$ \\
K2 & King, $W_0=8$ & 2.35 & 15 & high $W_0$ ($c_{\rm h}$) \\
P1 & Plummer, $c_{\rm h}=20$ & 2.35 & 15 & high $c_{\rm h}$ \\
\hline
\end{tabular}
\end{minipage}
\label{tab:ref}
\end{table*}

\subsubsection{Global behaviour}

\begin{figure}
\centerline{\epsfig{figure=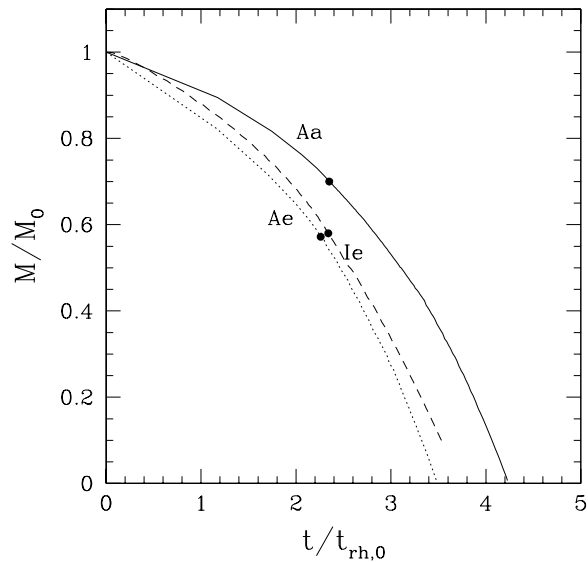, height=9cm}}
\caption{The evolution of mass of the model $R$ 
of three different assumptions ($Aa$, $Ae$, and $Ie$). 
The mass loss rate
is nearly a constant for single component models, but increases with time
in our multi-mass models. This is due to the fact that the mass function
becomes {\it flatter} as the cluster loses mass. The flatter mass function
means higher mass loss rate.}
\label{massfig}
\end{figure}

The evolution of the total mass is shown in Fig. \ref{massfig} 
with different criteria for ejection of stars. Unlike single
component models, the mass loss rate is not a constant, but increases 
with time. The increase of ${\dot M}$ is due to the fact that the
mass function becomes flatter as the cluster loses mass. (The mass loss rate
becomes larger for clusters with flatter mass function.)
As discussed in \S 3, the apocenter criterion ($Aa$) gives longer lifetime
than the energy criterion ($Ae$), but it is not a 
significant effect. For comparison, we have also shown the mass evolution
for the isotropic model ($Ie$) with the same initial conditions. 
The time of maximum
collapse is shown as dots in this figure.

\subsubsection{Degree of anisotropy}

\begin{figure}
\centerline{\epsfig{figure=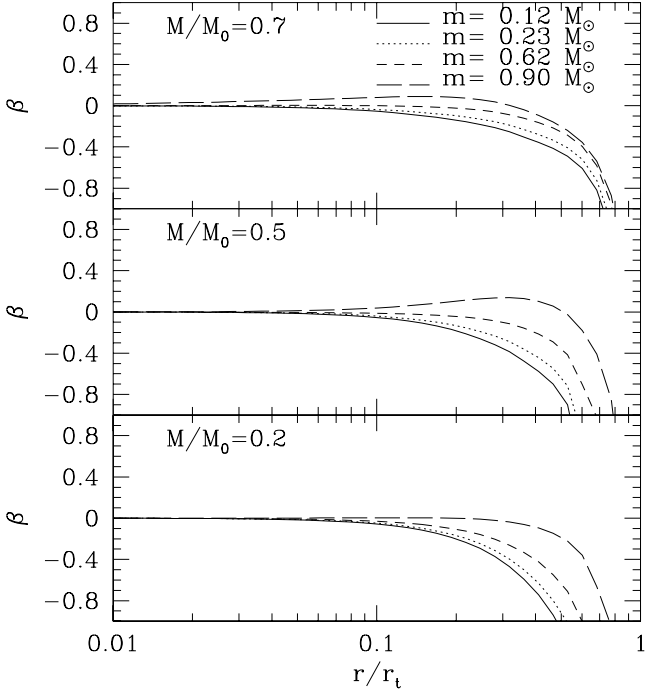, height=10cm}}
\caption{The dependence of $\beta$ on $r$ at three different epochs:
$M/M_0$=0.7, 0.5 and 0.2 for $R$ models with the apocenter condition. 
The core-collapse occurs when $M/M_0\approx$ 0.70. Note that the low mass 
components show more tangential anisotropy than high mass stars. This
is opposite to the isolated cluster models.
}
\label{aniso-apo}
\end{figure}

\begin{figure}
\centerline{\epsfig{figure=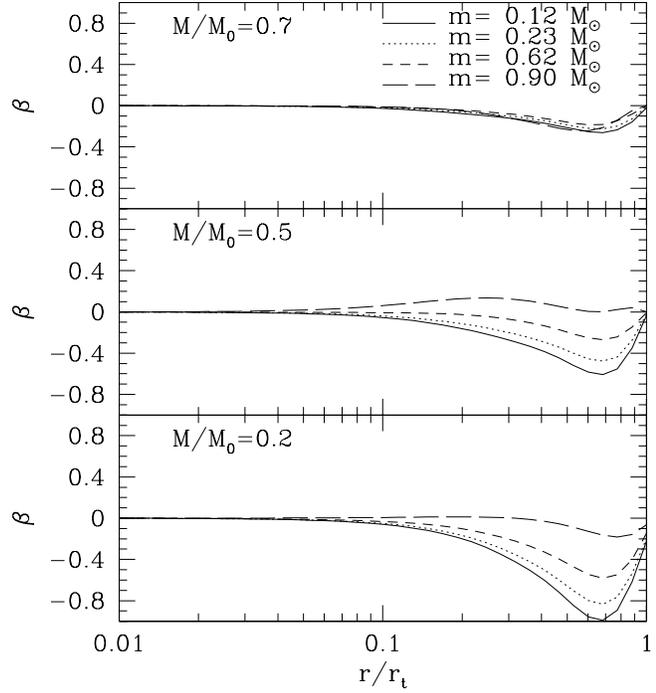, height=10cm}}
\caption{The dependence of $\beta$ on $r$ at three different epochs:
$M/M_0$=0.7, 0.5 and 0.2 for $R$ models with energy condition.
Note that the core-collapse occurs when $M/M_0\approx 0.57$. 
The velocity distribution is forced to become isotropic at
tidal boundary in the energy condition.
}
\label{aniso-en}
\end{figure}

The behaviours of the degree of velocity 
anisotropy $\beta$ for the apocenter criterion
are shown in Figure \ref{aniso-apo}. The 
upper, middle and lower panels show $\beta (r)$ at the time 
$M/M_0$=0.7, 0.5 and 0.2, respectively.
The same figure as Fig. \ref{aniso-apo} 
for the energy criterion is shown in Fig. \ref{aniso-en}. 

Note that the sign of $\beta$ can be both positive (radial anisotropy) and 
negative (tangential anisotropy). 
If a cluster is isolated,
strong radial anisotropy develops in the halo for all components,
as shown by Takahashi (1997).
However the present model cluster is in a rather strong tidal field:
the initial King model with $W_0=3$ has
$c_{\rm h}=3.7$.
Therefore, from the discussion in \S\,\ref{sec:ch}, we may expect that
the development of such radial anisotropy is strongly suppressed 
in the present case.
From the profiles shown in Fig. \ref{aniso-apo} (\Aa\ model),
in fact, only weak radial anisotropy is seen for the high mass components
at intermediate radii.
The low mass components tend to have tangential anisotropy at all radii.
A similar trend is seen in Fig. \ref{aniso-en} (\Ae\ model).
This is in sharp contrast with isolated clusters where strong 
radial anisotropy develops among low mass components (Takahashi 1997).
In the top panel of Fig. \ref{aniso-en}, the high mass components also 
have tangential anisotropy. This is due to the initial cooling of
the high mass components (Takahashi 1997).
The velocity distribution becomes mostly
tangential near the tidal boundary regardless of the mass. The tangential
anisotropy was also noted by Oh \& Lin (1992), and Kim \& Oh (1999)
from their hybrid
calculation of N-body and Fokker-Planck.

The behaviour of $\beta (r)$ becomes different for apocenter and energy 
criteria in the outer parts, because the energy criterion forces 
$\beta$ at tidal boundary to become zero (i.e., isotropic).

\subsubsection{Density profiles}

\begin{figure*}
\centerline{\epsfig{figure=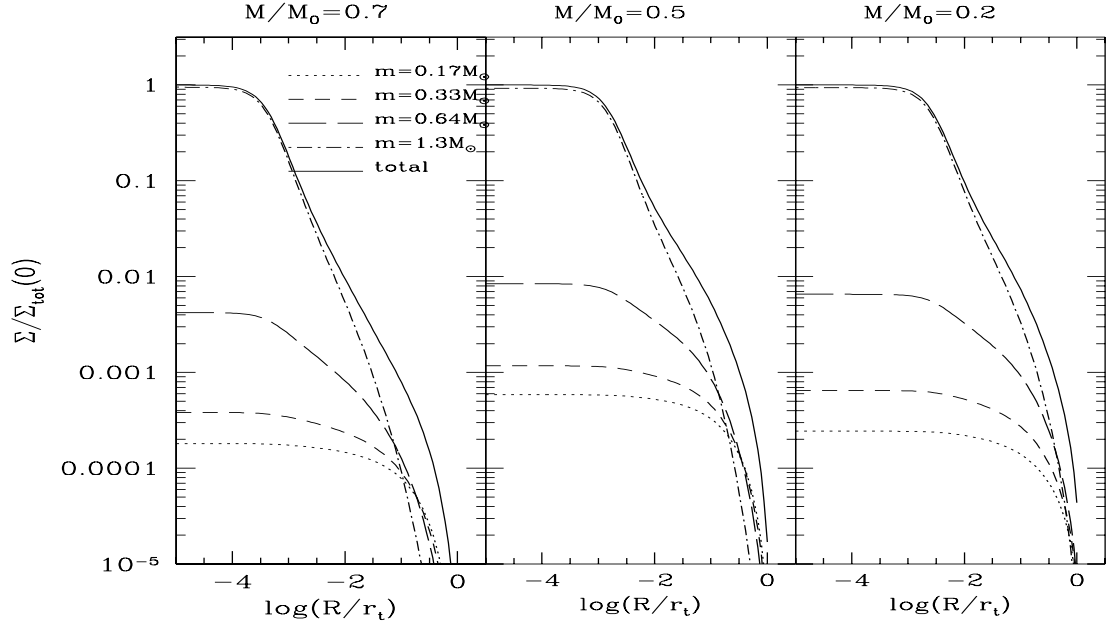, height=10cm,width=15cm}}
\caption{Surface density distribution of individual mass components at three
epochs for the $R$ model  with apocenter criterion ($Aa$).}
\label{density_a}
\end{figure*}

\begin{figure*}
\centerline{\epsfig{figure=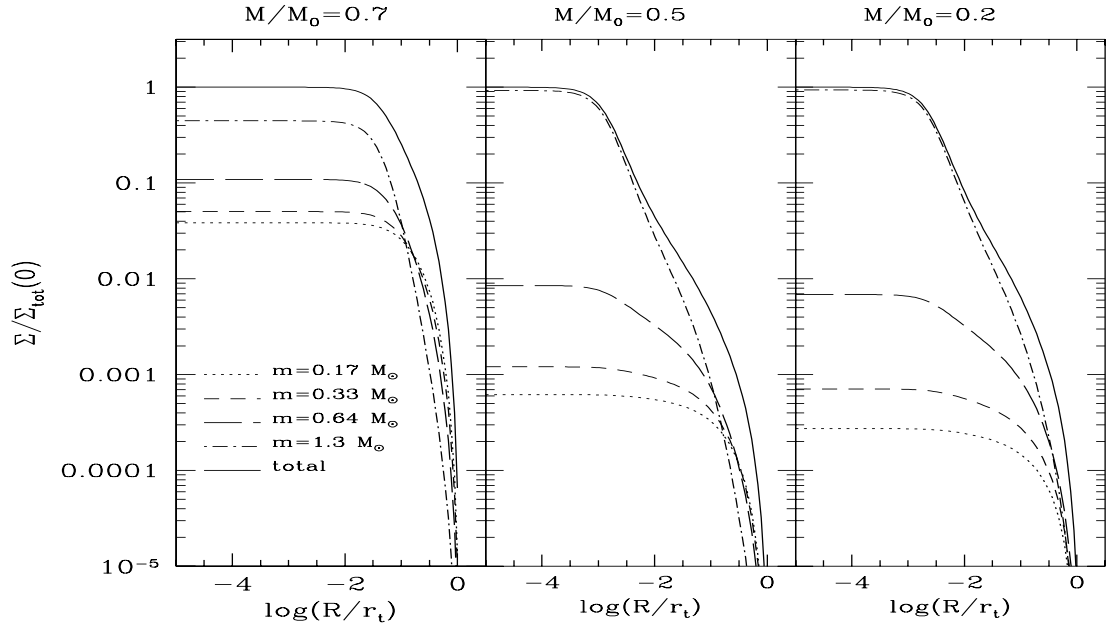, height=10cm,width=15cm}}
\caption{Surface density distribution of individual mass components at three
epochs for the $R$ model  with energy criterion ($Ae$).}
\label{density_e}
\end{figure*}

The profiles of surface densities for different mass components are shown in 
Fig. \ref{density_a} for a model with the apocenter condition at the epochs
of $M/M_0=0.7$, 0.5 and 0.2. Fig. \ref{density_e} shows
the density profiles for a model with the energy condition. The density
profiles at $M/M_0$=0.5 and 0.2 are similar for the apocenter
condition and the energy condition. The big difference at $M/M_0$=0.7
is purely due to the fact that the core-collapse occurs when $M/M_0\approx$ 0.7
for the \Aa\ model while it occurs somewhat later for the \Ae\ model.
Therefore, the difference in the degree of anisotropy
at outer radii does not play important role in determining the
density distribution.

\begin{figure*}
\centerline{\epsfig{figure=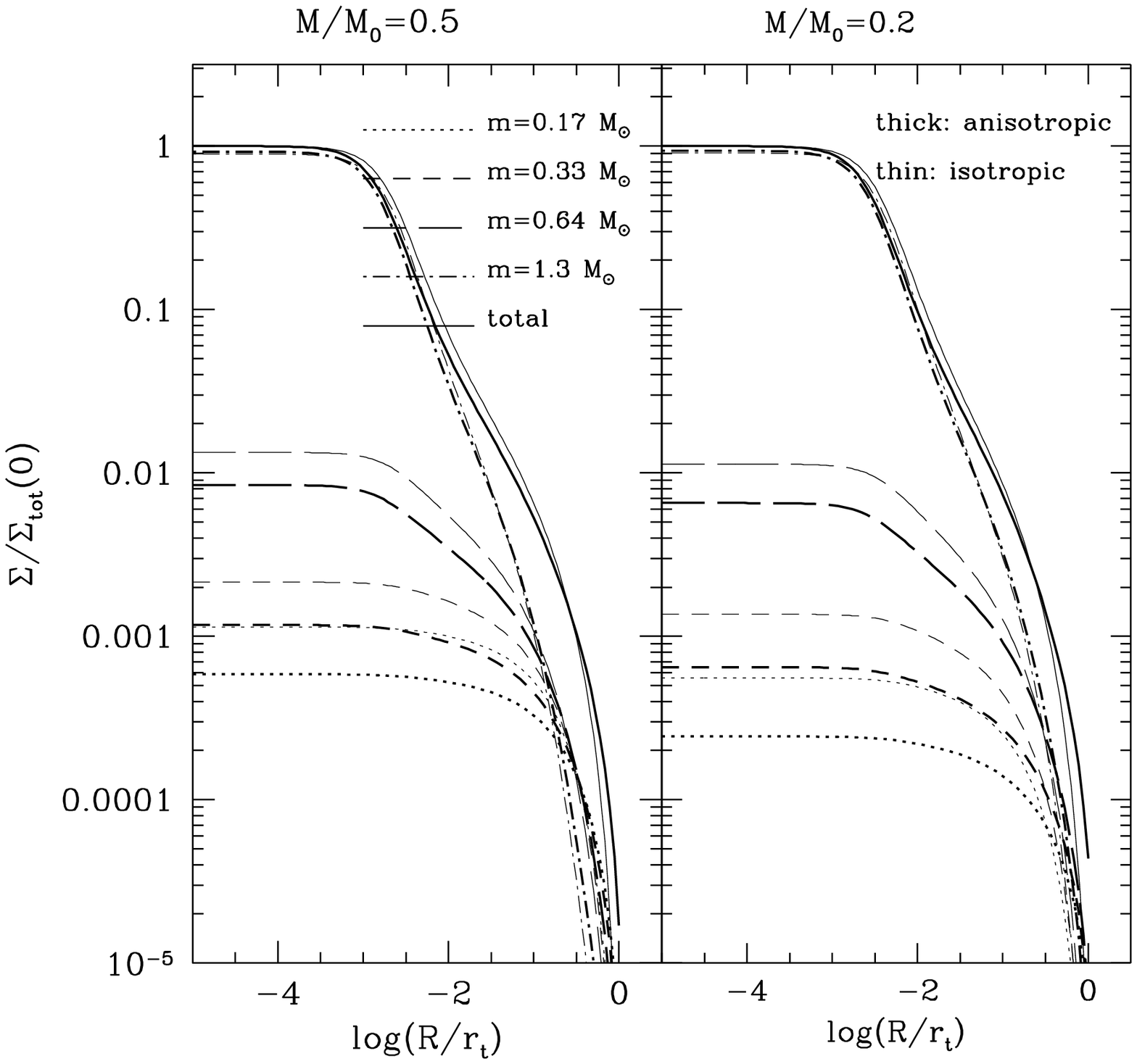, height=10cm,width=15cm}}
\caption{Comparison of density profiles of isotropic ($Ie$: thin lines) and
anisotropic ($Aa$: thick lines) Fokker-Planck results at $M/M_0$=0.5 and 0.2.}
\label{compare1}
\end{figure*}

The profiles at $M/M_0$=0.5 and 0.8 are compared with the isotropic model
in Fig. \ref{compare1}.
The radial profiles of the total density of isotropic and anisotropic models
have very little difference, but the detailed distribution of individual
mass components could be significantly different for these two models.
From these comparisons, we expect that there
will be some difference in the behaviour of the mass function in the lower
mass parts at late phase of the evolution.

\begin{figure}
\centerline{\epsfig{figure=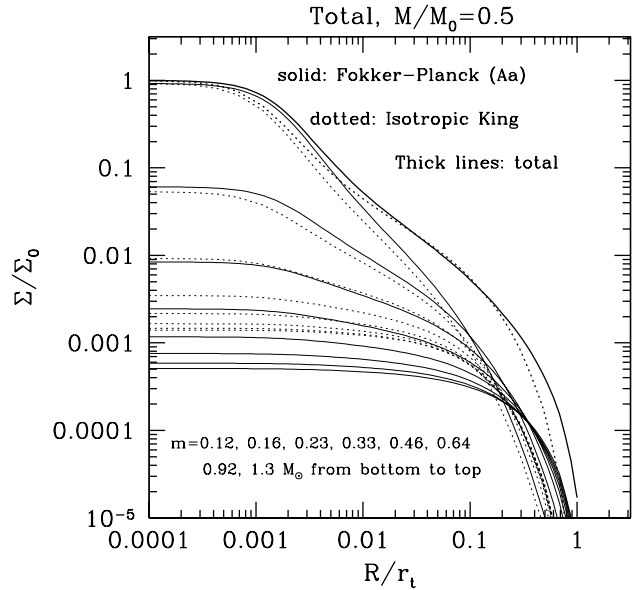, height=10cm}}
\caption{Comparison of the surface density profiles of the \Aa\ Fokker-Planck
model at $M/M_0$=0.5 and the best-fitting multi-mass King model.}
\label{compare2a}
\end{figure}

In Fig. \ref{compare2a}, 
we tried to compare the surface density profiles of the $R/Aa$ model at 
$M/M_0$=0.5 with 
the best-fitting multi-mass King models having the same mass function. 
Multi-mass King models have the following distribution 
function (e.g., Binney \& Tremaine 1987):
\begin{equation}
f_i (E) = C_i \left[ \exp( -E/\sigma_i^2) -\exp( -E_{\rm t}/\sigma_i^2) \right], \quad (i=1, 2, ..., K),
\end{equation}
where $\sigma_i$ satisfies the relation $m_i\sigma_i^2 = m_j\sigma_j^2$ for all 
$i$ and $j$. The constants $C_i$ should be determined by the given mass 
function. As in Fig. \ref{compare1}, the profiles of density of low mass 
stars do not fit King models: if the distribution of 
combined density is to be reproduced, the central densities of low mass
components of King model are higher than that generated by Fokker-Planck
results. This means that the lower mass stars are more
distributed at the outer parts than expected from King models. 
The mass segregation effect is greater than that predicted by multi-mass
King models. This is the same trend that was seen from the comparisons between
anisotropic and isotropic models. 

\begin{figure}
\centerline{\epsfig{figure=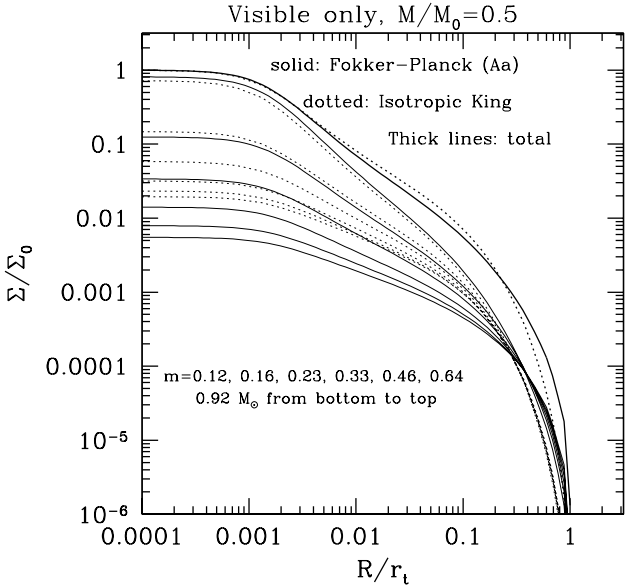, height=10cm}}
\caption{Same as Fig. \ref{compare2a}, but we have attempted to fit only 
the {\it visible} components, assuming that the highest mass component is
compact stars that have no luminosity. 
}
\label{compare2b}
\end{figure}

For the purpose of application to more realistic cases, we also
attempted to fit the surface density profiles of {\it visible} components
only, assuming that the highest mass bin ($m=1.3 \msun$) 
represents the remnant stars
such as neutron stars or massive white dwarfs. This is still far from
more realistic situation because the extension of simple power-law to 
massive remnant will give results in too much contribution from these stars.
Nonetheless, such a comparison could be worthwhile in understanding
the observed data. The result is shown in Fig. \ref{compare2b}. The fitting
model is the same as shown in Fig. \ref{compare2a}, but the difference between
the King model and the Fokker-Planck model is more pronounced. Choosing other
parameters for King models would not help to improve the fitting: the
difference simply is due to the fact that the distribution of stars are
substantially different between King models and Fokker-Planck results.
Because of stronger mass segregation in Fokker-Planck results, the fitting
of profiles in the intermediate region gives significant
differences in outer parts.

The fitting does not improve when one attempts to use anisotropic King
models (King-Michie models), because the King-Michie models always 
have $radial$ anisotropy regardless of mass, while 
the Fokker-Planck results give both radial and tangential anisotropy 
among different mass bins. 
The isotropic King models are usually better in reproducing the Fokker-Planck 
results. This is somewhat contradictory with the single component models:
during the early phase of the evolution, King-Michie models better
represent the anisotropic Fokker-Planck results of TLI than isotropic
King models.

\subsection{High $c_{\rm h}$ cases}

As described in \S\,\ref{sec:ch} for single-mass models, the development
of radial velocity anisotropy in the outer halo is strongly suppressed
in low half-mass concentration ($c_{\rm h}$) clusters.
To confirm this for multi-mass clusters we now examine the 
results for different
initial $c_{\rm h}$.

\begin{figure}
\centerline{\epsfig{figure=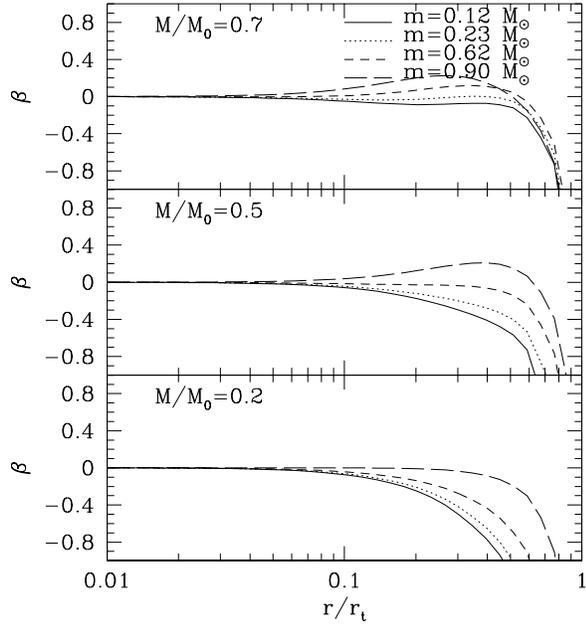, height=9cm}}
\caption{Same as Fig. \ref{aniso-apo} but for a model for
the initial conditions of the King model with $W_0=8$ (i.e., $K1$
which have  $c_{\rm h}=9.6$),
$\alpha=2.35$, and $\mu=15$.
The core-collapse occurs when $M/M_0$=0.99.
}
\label{aniso-apo-king8}
\end{figure}

Fig. \ref{aniso-apo-king8} shows the radial profiles of anisotropy $\beta$
at selected epochs for a reference initial condition except for 
$W_0=8$.
Initially this model has a larger $c_{\rm h}(=9.6)$ than the model discussed
in \S~\ref{sec:heggie} ($c_{\rm h}=3.7$).
Comparing Figs. \ref{aniso-apo} and \ref{aniso-apo-king8},
however, we find that
the behaviours of anisotropy are not very different between the two
models. For example,
the maximum value of $\beta$ is about 0.3 at the epoch of $M/M_0=0.5$
for both models.  This is not surprising 
because we may well expect that $c_{\rm h}$ for the King model of $W_0=8$
is still not large enough for strong radial anisotropy to develop
($c_{\rm h}<10$),
if we remember the results presented in \S~\ref{sec:ch}.

\begin{figure}
\centerline{\epsfig{figure=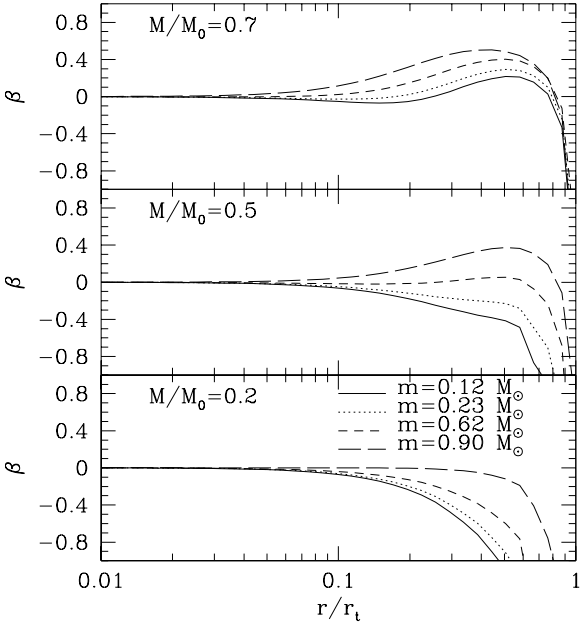, height=9cm}}
\caption{Same as Fig. \ref{aniso-apo} but for a model for
the initial conditions of the Plummer model (model $P1$)
with $c_{\rm h}=20$,
$\alpha=2.35$, and $\mu=15$.
The core-collapse occurs when $M/M_0$=0.99. 
}
\label{aniso-apo-plummer20}
\end{figure}

Fig. \ref{aniso-apo-plummer20} shows $\beta$ profiles
for an \Aa\ model for
the initial conditions of a Plummer model with $c_{\rm h}=20$,
$\alpha=2.35$, and $\mu=15$ (model $P1$).
In this case, as expected, stronger radial anisotropy 
(maximum $\beta \simeq 0.5$)
appears and is kept for a longer time than in the previous two cases.

\section{The Mass Function}\label{sec:mf}

It is observationally challenging task to obtain the mass function 
of globular clusters. Globular clusters are known as oldest Galactic 
objects and should possess very important memory of the early phase of 
the universe. For example, the IMF of globular clusters could provide a 
glimpse of the star formation process during the phase of very low heavy 
elements. 

\subsection{Evolution of The Mass Function}

\begin{figure}
\centerline{\epsfig{figure=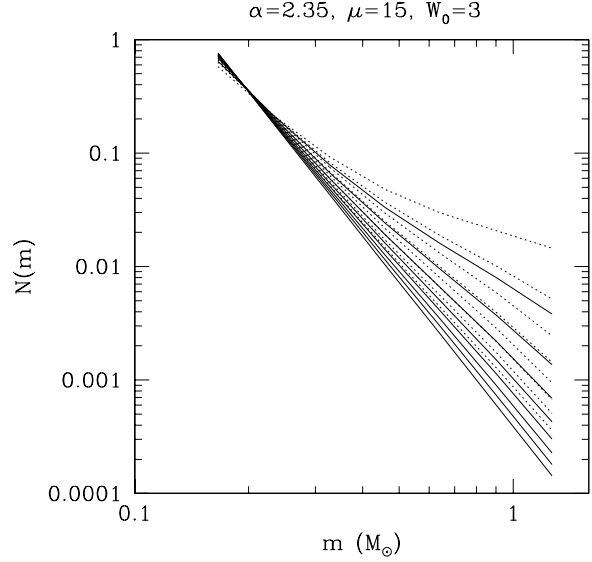, height=9cm}}
\caption{The mass function for $R/Aa$ model at
epochs of $M/M_0$=0.8, 0.7,
0.6, 0,5 0.4, 0.3, 0.2 and 0.1 (from steep to flat lines). 
The solid lines represent
the {\it global} mass function and the dotted lines are for
the mass functions within the half-mass radius.}
\label{mfheggie}
\end{figure}

The loss of mass occurs near the tidal boundary. The high mass stars 
gradually spiral into the central parts via dynamical friction and 
low mass stars are preferentially removed from the cluster. 
Therefore the shape of the mass function changes with time. This 
phenomenon has been observed from many numerical simulations (e. g., 
Chernoff \& Weinberg 1990; Lee, Fahlman \& Richer 1991; Lee \& Goodman 
1995; Vesperini \& Heggie 1997). The evolution of the 
mass function at several different epochs are shown in 
Fig. \ref{mfheggie}. 
The epochs are chosen according to the mass of the cluster: 
$M(t)/M_0$=0.8, 0.7, 0.6, 0.5, 0.4, 0.3, 0.2 and 0.1. 
This figure clearly shows the flattening of 
the mass function as the cluster loses mass. 
Also shown in this figure are the 
mass functions within the half-mass radius $r_h$ as dotted lines.

The shape of the mass function deviates from the simple power-laws
as a result of the selective mass loss: the average slope of the
mass function
decreases with time, and the slope of the mass function is
steeper at lower mass than at higher mass. Although the mass loss
rate is larger for the lower mass, it is not sensitive enough to 
flatten toward the lower mass. 
Such a behaviour (i.e., steeper slope at lower mass parts) 
appears to be inconsistent with the observational data 
for a number of globular clusters (e.g., Chabrier \& M\'era 1997).
The initial mass function of the globular clusters should not be
simple power-laws since the dynamical evolution only makes the situation
worse, but it is likely to flatyten (or even decreasing) toward the
lower mass.

\begin{figure}
\centerline{\epsfig{figure=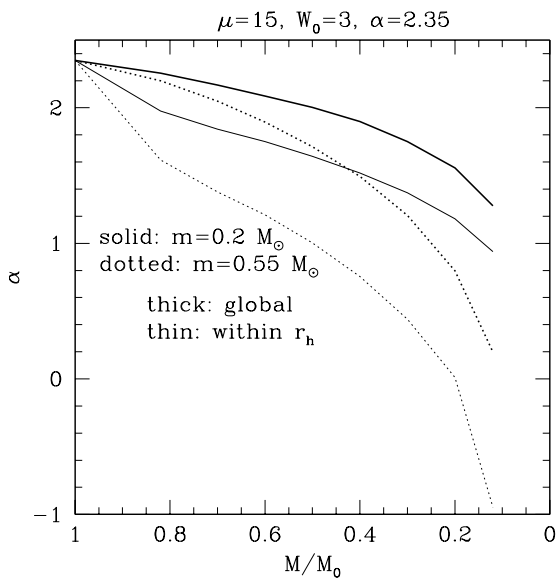, height=9cm}}
\caption{The evolution of power-law indices
of the global (thick lines) and
half-mass radius mass function for $R/Aa$ model, as a function of
cluster mass. The indices are computed for low mass
(m=0.2$\msun$: solid lines) and intermediate mass (m=0.55$\msun$:
dotted lines). The mass function is less steep at higher mass than
at lower mass.}
\label{alpha_hg}
\end{figure}

The evolution of mass function as a function of time can be measured by 
the variation of the power-law index. Suppose that the mass function can 
be approximated as a power-law with index $\alpha$ as shown in equation 
(\ref{eq:pmf}). 
As seen from Fig. \ref{mfheggie}, it is clear that the mass 
function does not follow a single power law as the cluster loses mass. 
Thus we have computed instantaneous $\alpha$ at two different
masses: $m=0.2\msun$ and 0.55 $\msun$. The
behaviour of these indices is shown in Fig \ref{alpha_hg} as a function of
$M(t)/M(0)$. The solid lines
represent $\alpha (0.2\msun$) and the dotted lines are for $\alpha
(0.55\msun$). Since the mass function depends on the location as well,
we have computed $\alpha$'s within the entire cluster (thick lines) and
within the half-mass radius (thin lines).

\begin{figure}
\centerline{\epsfig{figure=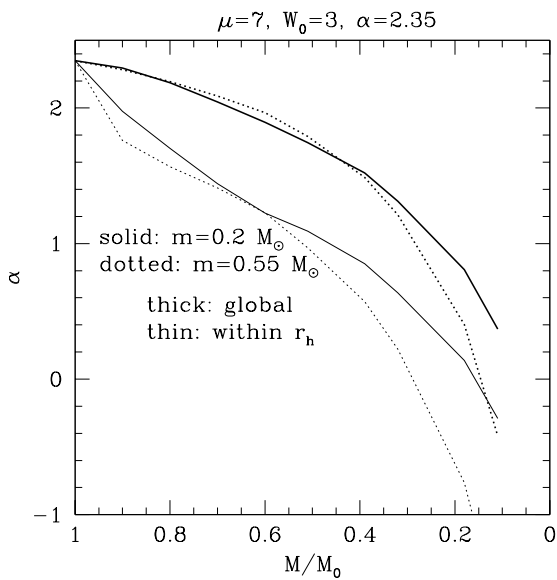, height=9cm}}
\caption{Same as Fig. \ref{alpha_hg} except for $\mu =7$ (i.e., model
$K1/Aa$).
Note that the difference between power indices at 0.2 and 0.55$\msun$ are
small compared to the $\mu = 15$ case.}
\label{alpha_low}
\end{figure}

The trends toward the flatter mass function with time are
clearly shown in this figure. The global mass function changes 
slowly in the early phase of the evolution, and rather abruptly in the
late phase. The mass function within the half-mass radius changes more
rapidly than the global mass function. 
The model with smaller $\mu$ gives more rapid variation in $\alpha$ with
time than that with higher $\mu$, as shown in Figs. \ref{alpha_hg} and
\ref{alpha_low}. We also note here that
the behaviour of $\alpha$ for the isotropic Fokker-Planck model
is very close to that of the anisotropic model. 

The variation of mass function observed by Richer et al. (1991)
could be most attributed by the dynamical evolution since the
mass function tends to be {\it flatter} for clusters with short
relaxation time. Clusters with short relaxation time are likely to have
lost significant fraction of the initial mass, mostly in the form
of low mass stars. Richer et al (1991) claim that there exists a
tight relation between the disruption time $T_D$ and the power-law index 
of the mass function. Although the amount of uncertainties is large,
the trend of flattening of mass function for clusters with significant
mass loss appears to be real.  

\subsection{The Relation between Local and Global MF}

The accurate photometry over any entire globular cluster is very
difficult. The luminosity function (and thus mass function) is usually
measured in limited range of the cluster. Since the mass function varies
with the location if the cluster has undergone significant evolution, it 
is important to understand the relationship between the locally determined
mass function and the `global' mass function.

\begin{figure}
\centerline{\epsfig{figure=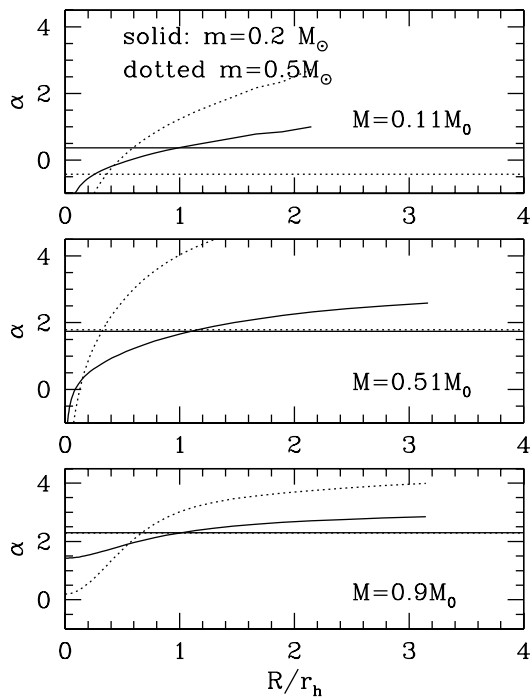, height=10cm}}
\caption{The variation of power-indices of mass functions over
the {\it projected} radius for three different epochs: 
M=0.9, 0.51 and 0.11 $M_0$
for  $\mu =7$ model with Salpeter IMF (i.e., model $K1/Aa$). 
The solid and dotted lines
represent for $m$=0.2 and 0.55 $\msun$, respectively. The horizontal
lines are the indices for the global mass function.}
\label{alpha_s}
\end{figure}

We have shown the variation of $\alpha$ as a function of projected radius 
for three different epochs in Fig. \ref{alpha_s} for 0.2$\msun$ (dotted lines)
and 0.55$\msun$ (solid lines). 
Also shown as horizontal lines are the power-law indices of the global mass 
function at that epoch. 
We have seen from Fig. \ref{alpha_low} that the global mass functions 
measured at 0.2 and 0.55 $\msun$ are rather similar. However, the
local mass function changes by a great amount. The 0.2$\msun$
mass function varies slowly over radius, but 0.55 $\msun$ one
varies a lot. This is due to the fact that the high mass stars are
concentrated toward the central parts. Thus one has to be  very careful
in determining the mass function from the observations. The power-law index
of the global mass function at 0.2$\msun$ roughly coincides with the 
local value near
the half-mass radius. The mass function near the half-mass radius
appears to represent the global mass function,  when measured at the
lower mass. This is also seen in N-body calculations by Vesperini \& 
Heggie (1997). 
The power-law index at $r_h$ for 0.55 $\msun$, however, deviates 
significantly from the global value. Therefore, the determination of
global mass function remains observationally challenging task.

\begin{figure*}
\centerline{\epsfig{figure=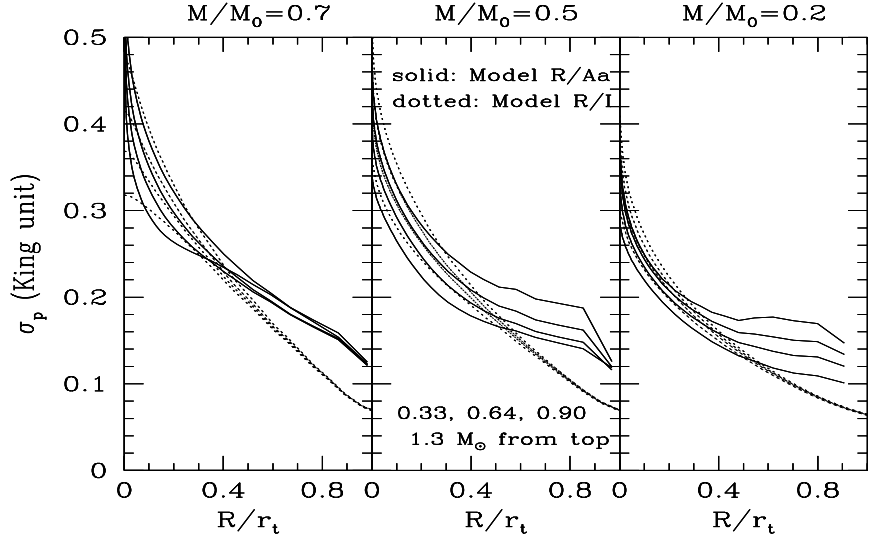, height=9cm, width=12cm}}
\caption{Projected velocity dispersion profiles for $R/Aa$ and $R/Ie$ models
at three different epochs: $M/M_0=0.7, ~ 0.5,$ and 0.2. The post
collapse expansion begins at near $M/M_0 \approx 0.7$.
Note that the velocity profiles are much flatter for the anisotropic model
with apocenter criterion than the isotropic model.}
\label{vproj}
\end{figure*}

\section{Velocity Profiles}\label{sec:vd}
So far we have examined density profiles and mass functions. 
We should be able to obtain more detailed dynamical information by
measuring the velocity dispersion profiles of the stars in the cluster.
In principle, both projected and transverse velocities can be
measured if the astrometric accuracy becomes of order of $10^{-6}$
arcseconds, as the astrometric project of GAIA is targeting. However,
only the projected velocity information is available for most of the
clusters in the Galaxy. We now discuss the prediction of projected
velocity dispersions of the cluster stars by our Fokker-Planck models.

In Fig. \ref{vproj}, we have plotted the projected velocity dispersion
profiles for the model $R/Aa$ at $M/M_0$=0.7, 0.5 and 0.2 for four different 
mass components,
together with the corresponding $Ie$ at 
the same epoch (measured by $M/M_0$). The maximum collapse
occurs near $M/M_0$=0.7. The unit of velocity in this plot
is $\sqrt{GM_0/r_0}$. The behaviour of velocity profiles for $R/Ae$
model was not shown here, but we note that the velocity profiles of such
a model behave similarly to those of isotropic model. This may be due
to the similarity in tidal boundary condition.

The $Aa$ model predicts the velocity profiles {\it flatter} than the isotropic
model. Since the cluster masses and the tidal radii are kept constant 
for different models, the difference in velocity profiles is
related to the difference in cluster structure. For example, the $Aa$ model
generally exhibits lower central velocity dispersion than the isotropic
model. This means that the anisotropic model has {\it larger}
$r_h$ than the isotropic model. 

Near the tidal boundary, the projected velocity profiles 
of the model $Aa$ deviate significantly
from the isotropic model. There are even mild bumps at around 0.5 $r_t$ for
low mass stars in late phase (i.e., $M/M_0 =0.2$). 
Such a behaviour is only seen for low mass stars in
$Aa$ models. Closer examination of
tangential and radial velocity dispersions revealed that the bump is
caused by the tangential component. The radial component monotonically
decreases like the isotropic model. 
The bumps appear in almost all models with apocenter
criterion for low mass stars during late postcollapse phase, but not in
other models. Therefore, this phenomena is clearly related to the
boundary conditions. We note here that the detailed observations by

Drukier et al. (1998) has discovered nearly flat or slightly rising
velocity dispersions from M15 at the outer parts. This seems to be the
indication of $tangentially$ anisotropic velocity distribution in the
outer parts of this cluster. It would be very desirable to carry out 
velocity measurements for other clusters.

\section{Summary}\label{sec:summary}

We have studied the evolution of multi-mass star clusters in the Galactic 
tidal field including the effects of velocity anisotropy. The radial anisotropy 
developed by dynamical relaxation tends to be suppressed by the presence 
of the tidal boundary. 
Except for very early epochs,
high mass stars show more radial (larger $\beta$) anisotropy
than low mass stars in general.
As the cluster loses a large amount of mass from the tidal boundary,
$\beta$ decreases, i.e., anisotropy becomes more tangential for all
mass species.
As a result, at late epochs of the cluster's life,
low mass stars generally have tangential (negative $\beta$)
anisotropy throughout the cluster, while high mass stars show weak radial 
anisotropy in the inner parts and tangential anisotropy in the outer parts. 
The overall degree of anisotropy depends on the half-mass concentration
$c_{\rm h}$ of the cluster. Larger $c_{\rm h}$ clusters allow the
development of stronger radial anisotropy in the halo.
Depending on the treatment of the tidal boundary condition, the detailed 
behaviour of the degree of anisotropy over radius changes near the tidal 
radius. The apocenter criterion gives nearly tangential motion for the 
stars near the tidal radius but the energy criterion forces isotropy there. 

The density profiles of tidally limited clusters are compared with the King 
models. Because of complex behaviour of the degree of anisotropy, finding 
the best-fitting King models becomes very difficult. The isotropic King 
models tend to give flatter profiles for low mass 
components compared to the Fokker-Planck results when the overall density 
profiles are matched.
Introduction of radial anisotropy gives only
worse fits, because most of the anisotropy is {\it tangential}.

We discussed the adequacy of converting the observed local mass function 
to the global mass function assuming that the cluster can be approximated by 
isotropic multi-mass King models.  Because of the discrepancy between 
King models and Fokker-Planck results, this process gives somewhat 
erroneous results of mass function for low mass components. We also 
looked for the best place where the local mass function is close to the 
global mass function and found that the half-mass radius should be a 
reasonable place. 

The mass function inevitably changes with time because of the selective 
loss of mass. The power-law indices of the mass function is found to be 
well correlated with the fraction of mass evaporated {\it via tidal 
overflow}. The amount of fractional mass loss can be determined by 
observed value of $t_{rh}$ and assuming the constant escape probability 
and coeval formation of all globular clusters. The observational data for 
the mass function are found to be consistent with the notion that the
mass function is strongly influenced by the dynamical processes
(e.g., Richer et al. 1990; Piotto \& Zoccali 1999), but
IMF itself may have been quite different from cluster to cluster: some
clusters may have formed with very steep IMF.

The projected velocity profiles
of anisotropic models with apocenter criterion  
show significant deviation from isotropic models or anisotropic models 
with energy condition in the sense that the velocity profiles are
much flatter for $Aa$ models in the outer parts during the postcollapse
phase. This phenomena appears
to be consistent with the observational data for M15.

\section*{Acknowledgements}
This research was supported by the Research for the Future Program of
Japan Society for the Promotion of Science (JSPS-RFTP97P01102)
to KT and by KOSEF grant No. 
95-0702-01-01-3 to HML.

\label{lastpage}
\end{document}